\documentclass[prl,twocolumn,nofootinbib]{revtex4}
\usepackage{graphicx}
\usepackage{amssymb,amsmath,enumerate}

\newcommand{\be}{\begin{eqnarray}}
\newcommand{\ee}{\end{eqnarray}}
\newcommand{\D}{\mathrm{d}}

\begin{document}

\title{Adhesive loose packings of small particles}

\author{Wenwei Liu$^1$, Shuiqing Li$^1$\footnote{lishuiqing@tsinghua.edu.cn}, Adrian Baule$^2$, Hern\'an A. Makse$^3$\footnote{hmakse@lev.ccny.cuny.edu}}

\affiliation{ $^1$Key Laboratory for Thermal Science and Power
  Engineering of Ministry of Education, Department of Thermal
  Engineering, Tsinghua University, Beijing 100084, China\\ $^2$School
  of Mathematical Sciences, Queen Mary University of London, Mile End
  Road, London E1 4NS, UK\\ $^3$Levich Institute and Physics
  Department, City College of New York, New York 10031, USA }

\begin{abstract}

We explore adhesive loose packings of dry small spherical particles of
micrometer size using 3D discrete-element simulations with adhesive
contact mechanics. A dimensionless adhesion parameter ($Ad$)
successfully combines the effects of particle velocities, sizes and
the work of adhesion, identifying a universal regime of adhesive
packings for $Ad>1$. The structural properties of the packings in this
regime are well described by an ensemble approach based on a
coarse-grained volume function that includes correlations between bulk
and contact spheres. Our theoretical and numerical results predict:
{\it (i)} An equation of state for adhesive loose packings that
appears as a continuation from the frictionless random close packing
(RCP) point in the jamming phase diagram; {\it (ii)} The existence of
a {\it maximal loose packing} point at the coordination number $Z=2$
and packing fraction $\phi=1/2^{3}$. Our results highlight that
adhesion leads to a universal packing regime at packing fractions much
smaller than the random loose packing, which can be described within a
statistical mechanical framework. We present a general phase diagram
of jammed matter comprising frictionless, frictional, adhesive as well
as non-spherical particles, providing a classification of packings in
terms of their continuation from the spherical frictionless RCP.

\end{abstract}

\date{\today}

\maketitle

Jammed particle packings have been studied to understand the
microstructure and bulk properties of liquids, glasses and crystals
\cite{Bernal60,Parisi10} and frictional granular materials
\cite{coniglio,Andreotti13}. Two packing limits have been identified
for disordered uniform spheres: The random close packing (RCP) and
random loose packing (RLP) limits
\cite{Bernal60,Scott60,Onoda90,Ciamarra08,Jerkins08,Dong06,Song08,Farrell10}. The
upper RCP limit is reproduced for frictionless spheres at volume
fractions $\phi\approx 0.64$ and has been associated with a freezing
point of a 1st order phase transition
\cite{anikeenko,klumov,Jin10,Panaitescu12}, among other
interpretations \cite{Parisi10,torquato}. In the presence of friction,
packings reach lower volume fraction up to the RLP limit $\phi_{\rm
  RLP}\approx 0.55$ for mechanically stable packings
\cite{Onoda90,Jerkins08,Farrell10}. However, most packings of dry
small micrometer-sized particles in nature are not only subject to
friction, but also {\it adhesion} forces. In fact, van der Waals
forces generally dominate interactions between particles with
diameters of around $10 \mu m$ or smaller. In this case, the adhesive
forces begin to overcome the gravitational and elastic contact forces
acting on the particles and change macroscopic structural properties
\cite{Li11,Marshall14}. 

Despite the ubiquity of adhesive particle
packings in almost all areas of engineering, biology, agriculture and
physical sciences \cite{Marshall14,Dominik97,Blum04,Kinch06}, these
packings have so far not been systematically investigated. The
multi-coupling of adhesion, elastic contact forces and friction within
the short-range particle-particle interaction zone and their further
couplings with fluid forces (e.g., buoyancy, drag and lubrication)
across long-range scales make it highly difficult to single out the
effect of the adhesion forces alone. Previous studies have found
packing fractions of adhesive microparticles in a wide range of
$\phi=0.07-0.33$ for both uncompressed and compressed samples using a
random ballistic deposition technique \cite{Blum06}. A similarly broad
range of $\phi=0.23-0.41$ was found for particles with diameter of
$7.8-19.1\mu m$ using fluidized bed techniques \cite{Valverde04}.

In this letter, a prototypical packing system is introduced for the
simulation of random loose packings of {\it soft-sphere},
non-Brownian, adhesive particles with a discrete element method
(DEM). Here, the fluid effect is filtered out by assuming the
gravitational sediment under a vacuum condition. Most importantly, the
gravitational effect can be neglected when the system satisfies
$F_r=U_0/\sqrt{g H}>>1$, where $F_r$ is the {\it Froude} number (ratio
of inertia to gravity), $H$ the characteristic height of the
deposition control volume and $U_0$ the initial particle velocity at
the upper inlet boundary. For all runs in the numerical simulations,
we ensure that gravitational effects with respect to initial particle
inertia are less than $3\%$. Therefore, the adhesive packings simply
arise due to the competition between the particle inertia and
particle-particle interactions (e.g., adhesion, elasticity and
frictions). 

In a novel DEM framework specifically developed for adhesive grains
\cite{Li11, Marshall14}, both the transitional and rotational motions
of each particle in the system are considered on the basis of Newton's
second law (see Supplementary Information, Section I). The adhesive
contact forces $F_A$ include three terms, the normal adhesive contact
force $F_{\rm ne}$, the normal damping force $F_{nd}$ validated by
classic particle-surface impact experiments \cite{Li11,Marshall14},
and the tangential force due to the sliding friction. A JKR
(Johnson-Kendall-Roberts) model is employed to account for $F_{\rm
  ne}$ between the relatively compliant micro-particles, implying the
length scale of elastic deformation is large compared to the length
scale of the adhesive force (with the particles' Tabor parameter
larger than unity) \cite{Marshall14}. The dissipation terms, including
the sliding, twisting and rolling frictions in the presence of
adhesion, are all approximated by a linear spring--dashpot--slider
model with model parameters given in \cite{Yang}. The slider
considerations mean that the sliding, twisting and rolling resistances
all reach critical values, $F_{s,{\rm crit}}$, $M_{t,{\rm crit}}$ and
$M_{r,{\rm crit}}$, as three related displacements exceed the certain
limits. For displacements larger than those limits, the resistances
stay constant and the particles start to slide or spin. The critical
limits in presence of adhesion are given in the following equations
\cite{Marshall14,Dominik97,Heim99}, \be F_{s,{\rm
    crit}}&=&\mu_f|F_{\rm ne}+2F_C|\\ M_{t,{\rm crit}}&=&3\pi a
F_{s,{\rm crit}}/16\\ M_{r,{\rm crit}}&=&-4F_C(a/a_0)^{3/2}\Theta_{\rm
  crit}R, \ee where $\mu_f$ is the friction coefficient, $\Theta_{\rm
  crit}$ is the critical angle for the relative rolling of two
particles, and $F_C$ is the critical pull-off force expressed by the
{\it work of adhesion} (twice the surface energy, $w=2\gamma$):
$F_C=1.5\pi w R$. Here, $R$ is the effective radius between two
contacting particles ($1/R=1/r_{p_1}+1/r_{p_2}$), $a$ is the radius of
the contact area with $a_0$ at equilibrium in the JKR model. The
values or ranges of $\mu_f$, $\Theta_{\rm crit}$ and $F_C$ are
selected according to the data from atomic force microscopy
measurements \cite{Heim99,Jones04,Suemer08}.
 
\begin{figure}
\begin{center}
\includegraphics[width=7.5cm]{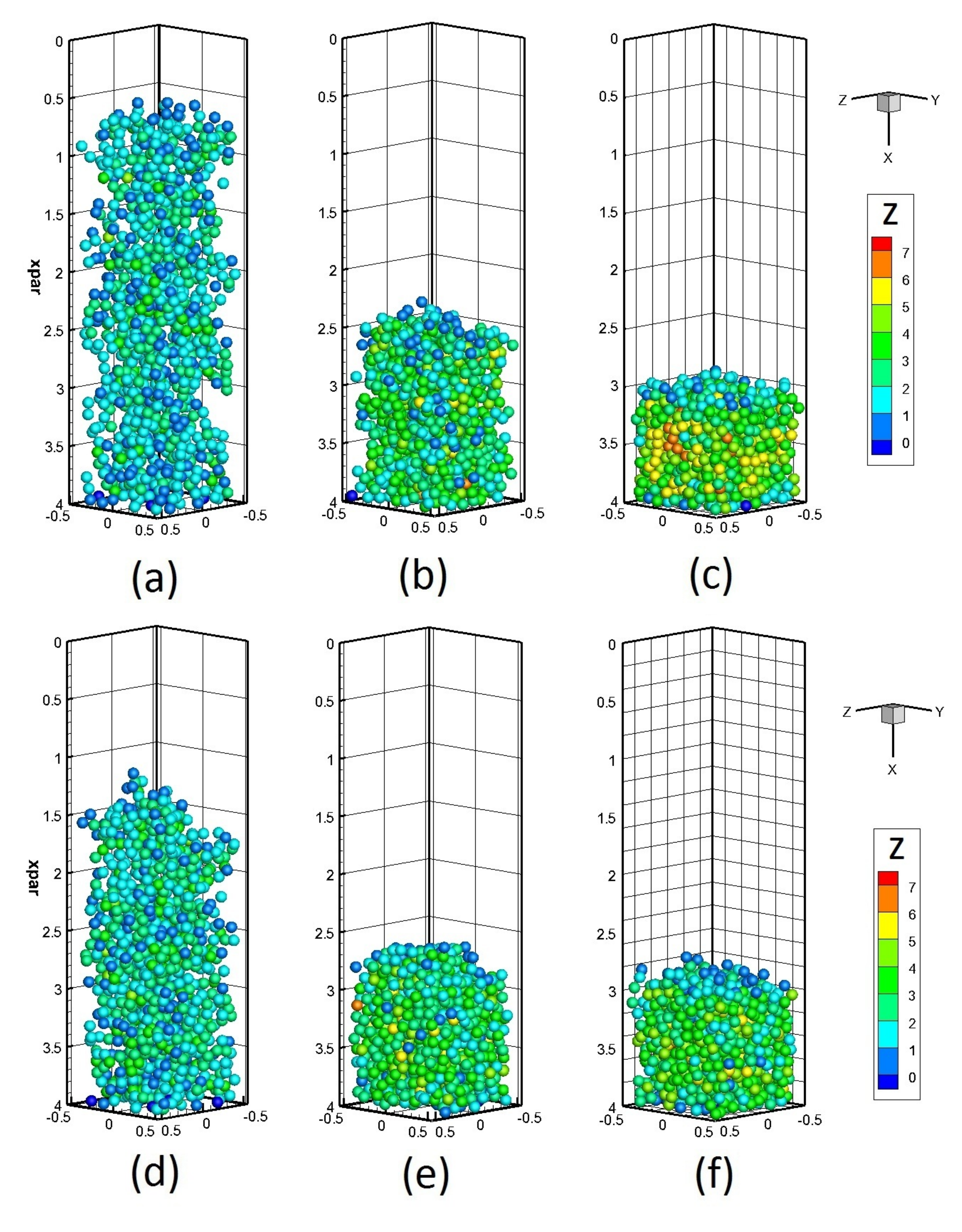}
\caption{\label{Fig_structure}(Colors online) Typical packing
  structure: different colors represent different coordination numbers
  $Z$. (a)(b)(c) stand for $U_0=0.5$, $2$ and $6m/s$, respectively
  with particle radius $r_p=1\mu m$ and work of adhesion
  $w=30mJ/m^2$. (d)(e)(f) stand for $r_p=1$, $5$ and $10\mu m$,
  respectively with $U_0=1m/s$ and $w=30mJ/m^2$.  }
\end{center}
\end{figure}

\begin{figure}
\begin{center}
\includegraphics[width=7.5cm]{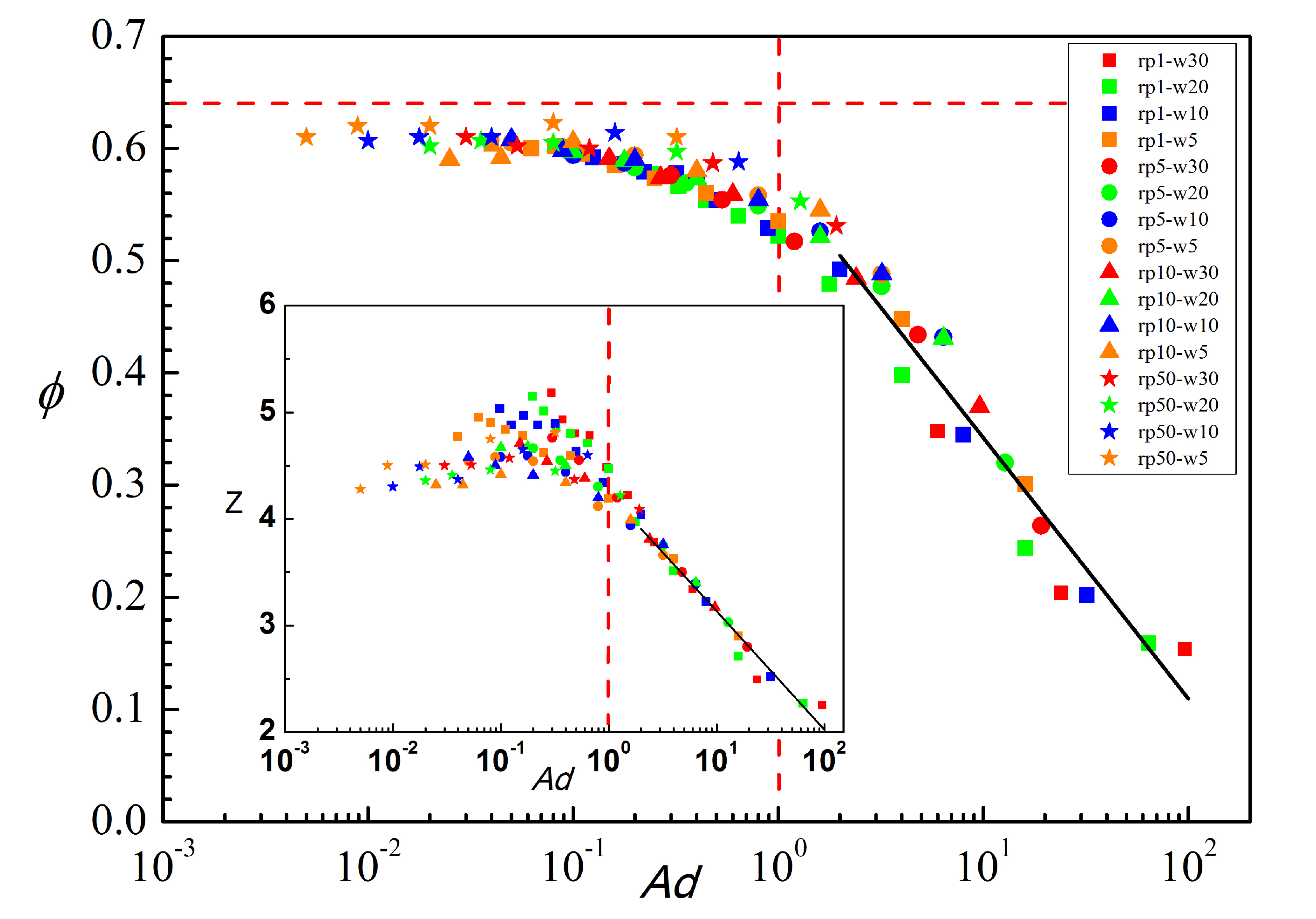}
\caption{\label{Fig_ad}(Colors online) Semi-log plot of the packing
  volume fraction as a function of adhesion parameter. The horizontal
  red dash line indicates the limitation of $\phi_{RCP}=0.64$ and the
  vertical one is the separation of $Ad=1$. The inset shows the
  variation of the coordination number $Z$ with $Ad$.  }
\end{center}
\end{figure}

The adhesive DEM simulation starts with the random free falling of
1,000 spheres with an initial velocity $U_0$ at a height $H$ (see the
Supplementary Information, Section II, for detailed physical and
geometrical parameters). The horizontal plane for particle deposition
has two equal edges of length $L$, with periodic boundary conditions
along the two horizontal directions. Here we focus on uniformly sized
particles, with particle radius ranging from $r_p=1$ to $50\mu m$.  A
sensitivity analysis shows that the difference in $\phi_{\rm RLP}$
between the cases $L=20r_p$ and $L=40r_p$ is negligible, indicating
the $L=20r_p$ is large enough to reproduce bulk properties. Then, we
set $L=20r_p$.  The work of adhesion, e.g., for silica microspheres is
reported at $20-30 mJ/m^2$ \cite{Marshall14,Heim99}. Setting
$w=30mJ/m^2$, the simulations indicate that both particle deposition
velocities and particle sizes significantly affect packing
structures. As shown in Fig.~\ref{Fig_structure} (more details in
supplementary Fig. S2), either large velocity and small size, or small
velocity and large size can produce a relatively dense packing. A
dimensionless adhesion parameter $Ad=w/(2\rho_pU_0^2r_p$), defined as
the ratio between interparticle adhesion and particle inertia by Li
and Marshall \cite{Li07}, can be used to quantify the combined effect
of velocity and size. Figure ~\ref{Fig_ad} shows the variation of
packing volume fraction as a function of $Ad$ for packings with
$r_p=1,5,10,50\mu m$ with $w$ ranging from $5-30 mJ/m^2$. In the case
of $Ad<1$, the data points are scattered between RLP ($0.55$) and RCP
($0.64$), since particle inertia dominates the adhesion and
frictions. However, when $Ad >1$, we obtain an adhesion-controlled
regime, in which the volume fraction decreases exponentially with
increasing $Ad$, becoming linear at large $Ad\sim10$. The lowest
packing density achieved is at $\phi=0.154$ when $Ad$ is as high as
$96$, which agrees well with the data from a random ballistic
deposition experiment \cite{Blum06}.

\begin{figure}
\begin{center}
\includegraphics[height=5.7cm]{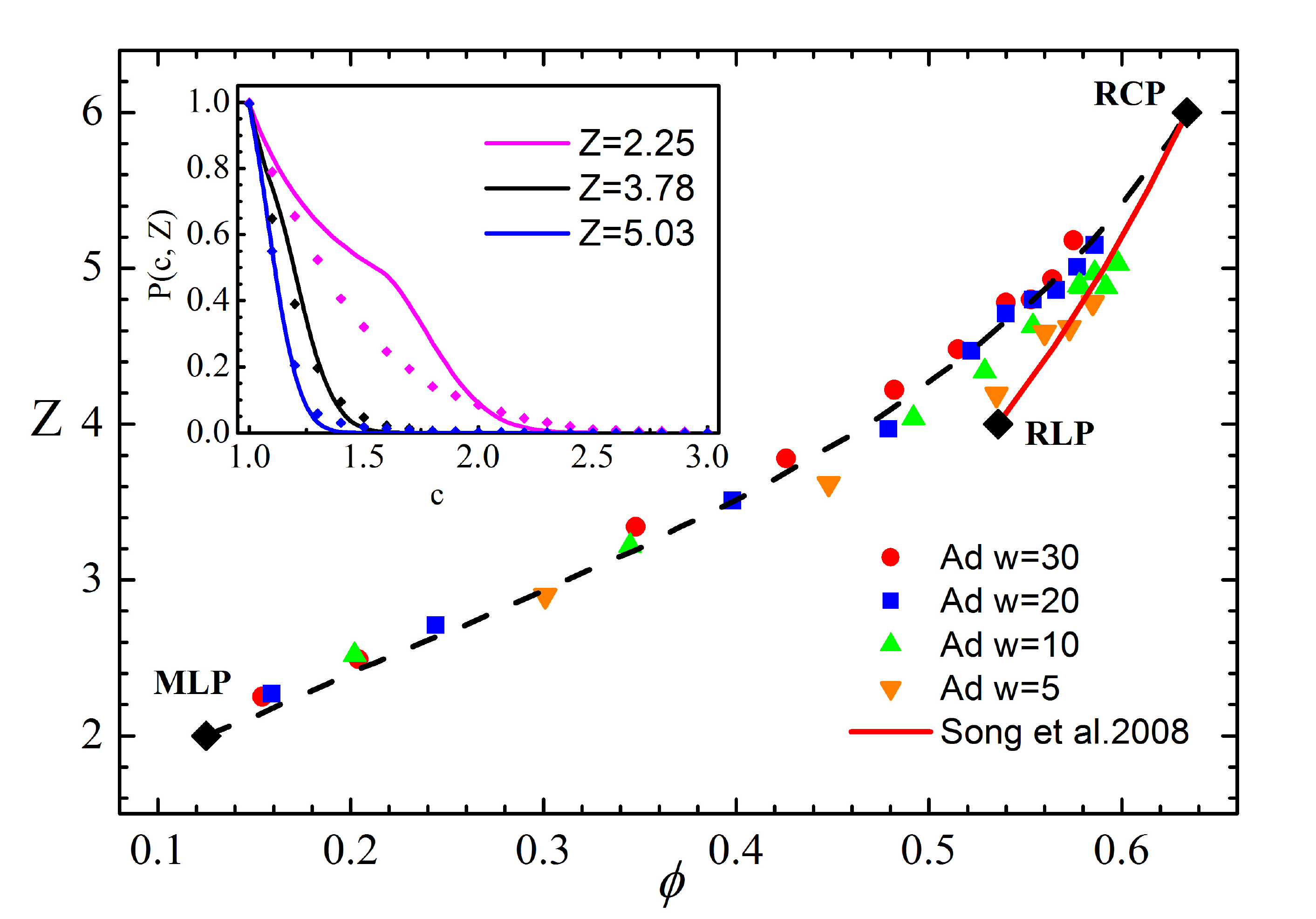}
\caption{\label{Fig_zphiad}(Colors online) Equation of state for adhesive packings: Simulation data and theoretical prediction from Eqs.~(\ref{wbar})--(\ref{g2}) with a single fitting parameter (black dashed line). The equation of state for frictional sphere packings from Ref.~\cite{Song08} is indicated. Inset: Comparison of the CDF P(c,Z) with simulation data (dots).
}
\end{center}
\end{figure}

In addition to $\phi$, a reproducible observable of the packing is the
coordination number $Z$, which denotes the average number of contacts
of a sphere in the packing. The isostatic conjecture predicts the
upper and lower bounds of $Z=2d_f$ and $Z=d_f+1$ for frictionless and
infinitely rough hard-spheres, respectively, with $d_f=3$ degrees of
freedom. In Fig.~\ref{Fig_ad} (inset) we see that for $Ad<1$ the
packings lie indeed within the isostatic limits reaching the
infinitely rough value $Z=4$ at $Ad\approx1$. This indicates that weak
adhesion has a similar effect on the packing as strong friction. For
$Ad>1$ the adhesive packings fall on a unique curve, analogous to the
$\phi$ dependence. The lowest $Z$ reached in our simulations is
$Z=2.25$. Combining our results for $\phi$ and $Z$ thus highlights a
universal adhesive regime characterized by the dimensionless parameter
$Ad$. The resulting curve in the $Z$-$\phi$ plane, parametrized by
$Ad$, can be considered as an equation of state of packings dominated
by adhesion (see Fig.~\ref{Fig_zphiad}).

We now derive an analytical representation of the adhesive equation of
state. To this end we introduce a statistical mechanical framework for
particles with adhesive interactions in the spirit of Edwards'
ensemble approach \cite{Song08,Jin10b,Baule13}. We start with the
Voronoi volume $W_i$ of a reference particle $i$, which provides a
tessellation of the total volume of the packing:
$V=\sum_{i=1}^NW_i$. The key step is to use a statistical mechanical
description, where we consider the average Voronoi volume
$\overline{W}=\left<W_i\right>$. This implies that $V=N \overline{W}$
and the packing fraction follows as $\phi=V_0/\overline{W}$. Here,
$V_0$ is the volume of a sphere with radius $r_p$ in the packing. In
turn, $\overline{W}$ can be expressed exactly in terms of the pdf
$p(c,Z)$ for finding the boundary of the Voronoi volume at a distance
$c$ from the sphere centre for a given $z$. We have
\cite{Song08,Baule13}

\be
\label{wbar}
\overline{W}=V_0+4\pi\int_{r_p}^\infty\D c\, c^2 P(c,Z), \ee where
$P(c,Z)$ is the CDF; $p(c,Z)=-\frac{\D}{\D c}P(c,Z)$. For $P(c,Z)$ one
can derive a Boltzmann-like form using a factorization assumption of
the multi-particle correlation function into pair correlations
\cite{Jin10b}

\be
\label{pc}
P(c,Z)=\exp\left\{-\rho\int_{\Omega(c)}\D
\mathbf{r}\,g_2(\mathbf{r},Z)\right\}.  \ee Here,
$\rho=N/V=1/\overline{W}$ is the number density and
$g_2(\mathbf{r},Z)$ the pair correlation function of two spheres
separated by $\mathbf{r}$. The volume $\Omega(c)$ is an excluded
volume for the $N-1$ spheres outside of the reference sphere, since
otherwise they would contribute a Voronoi boundary smaller than
$c$. Typical models for $g_2$ consist of a step function plus a
delta-peak to model the contribution of bulk and contacting particles
(with a given coordination number $Z$), respectively
\cite{Torquato06,Jin10b}:
$g_2(\mathbf{r},Z)=\Theta(r-2r_p)+\frac{Z}{\rho
  \lambda}\delta(r-2r_p)$, where $\lambda$ is an appropriate
constant. In order to describe the effect of correlations due to
adhesion, we assume a gap of width $b(Z)$ that models the increased
porosity at a given $Z$ compared with adhesion-less packings. This
suggests to express $g_2$ as \be
\label{g2}
g_2(\mathbf{r},Z)=\Theta(r-(2r_p+b(Z)))+\frac{Z}{\rho \lambda}\delta(r-2r_p).
\ee

Clearly, $b(Z)$ is a smoothly decreasing function, so that we can
assume, e.g., the simple parametric form
$b(Z)=c_1+c_2e^{-c_3Z}$. There are then two natural boundary
conditions that constrain $(c_1, c_2, c_3)$, the three parameters in
the theory: {\it (i)} At the isostatic limit $Z=6$, we expect to
recover the frictionless RCP value, since this value of $Z$ represents
a maximally dense disordered packing of spheres. We obtain from
Eqs.~(\ref{wbar})--(\ref{g2}) indeed the prediction of
Ref.~\cite{Song08} for RCP at $\phi=\sqrt{3}/(1+\sqrt{3}) = 0.634$
provided we have $b(6)=0$ and $\lambda$ assumes the value
$\lambda=16\pi r_p^2/\sqrt{3}$. Moreover, we need to account for low
dimensional corrections due to the hard-core excluded volume of the
reference sphere, such that $\rho\to
\overline{\rho}=1/(\overline{W}-V_0)$, where $V_0$ is the volume of a
sphere with radius $r_p$ \cite{Jin10b}. The constraint {\it (i)} thus
fixes $\rho$ and $\lambda$, as well as one of the parameters in
$b(Z)$, say $c_1$. {\it (ii)} In addition, we conjecture the existence
of a {\it maximally loose packing} (MLP) at $Z=2$ and $\phi=1/2^3$
which yields $b(2)=1.47$ and fixes a second parameter, $c_2$. This is
motivated by the fact that $\phi=1/2^d$ is the well-known lower bound
of saturated sphere packings in $d$ dimensions
\cite{Torquato06}. Moreover, $Z=2$ is the lowest possible value for a
physical packing: If $Z<2$ there are more spheres with a single
contact (i.e., dimers) than with three or more contacts.

Solving Eqs.~(\ref{wbar})--(\ref{g2}) numerically for $\overline{W}$
(and thus $\phi$) with the functional form of $b(Z)$ leads to a family
of curves with a single free parameter, $c_3$. Fitting this parameter
to the available data yields $c_3=1$. We then obtain a unique equation
of state $\phi(Z)$ for adhesive packings as shown in
Fig. \ref{Fig_zphiad} which agrees remarkably well with the simulation
data. Furthermore, the adhesive $\phi(Z)$ equation of state appears as
a continuation of the frictionless RCP at $(Z,\phi) = (6, 0.634).$ For
large $Ad$ values the MLP point at $(2,0.125)$ is indeed approached in
the $Z$-$\phi$ plane. A comparison of the theoretically obtained
$P(c,Z)$ with simulation data is shown in the inset of
Fig.~\ref{Fig_zphiad}. We observe that the agreement is good for small
$c$ values throughout the range of adhesive packings. Since this range
of $c$ values provides the dominant contribution to $\overline{W}$ in
Eq.~(\ref{wbar}), we conclude that our phenomenological approach
captures well the microstructure of adhesive packings in the first
coordination shell, which provides a good first moment from $P(c,Z)$
as compared with simulations.

\begin{figure}
\begin{center}
\includegraphics[height=5.7cm]{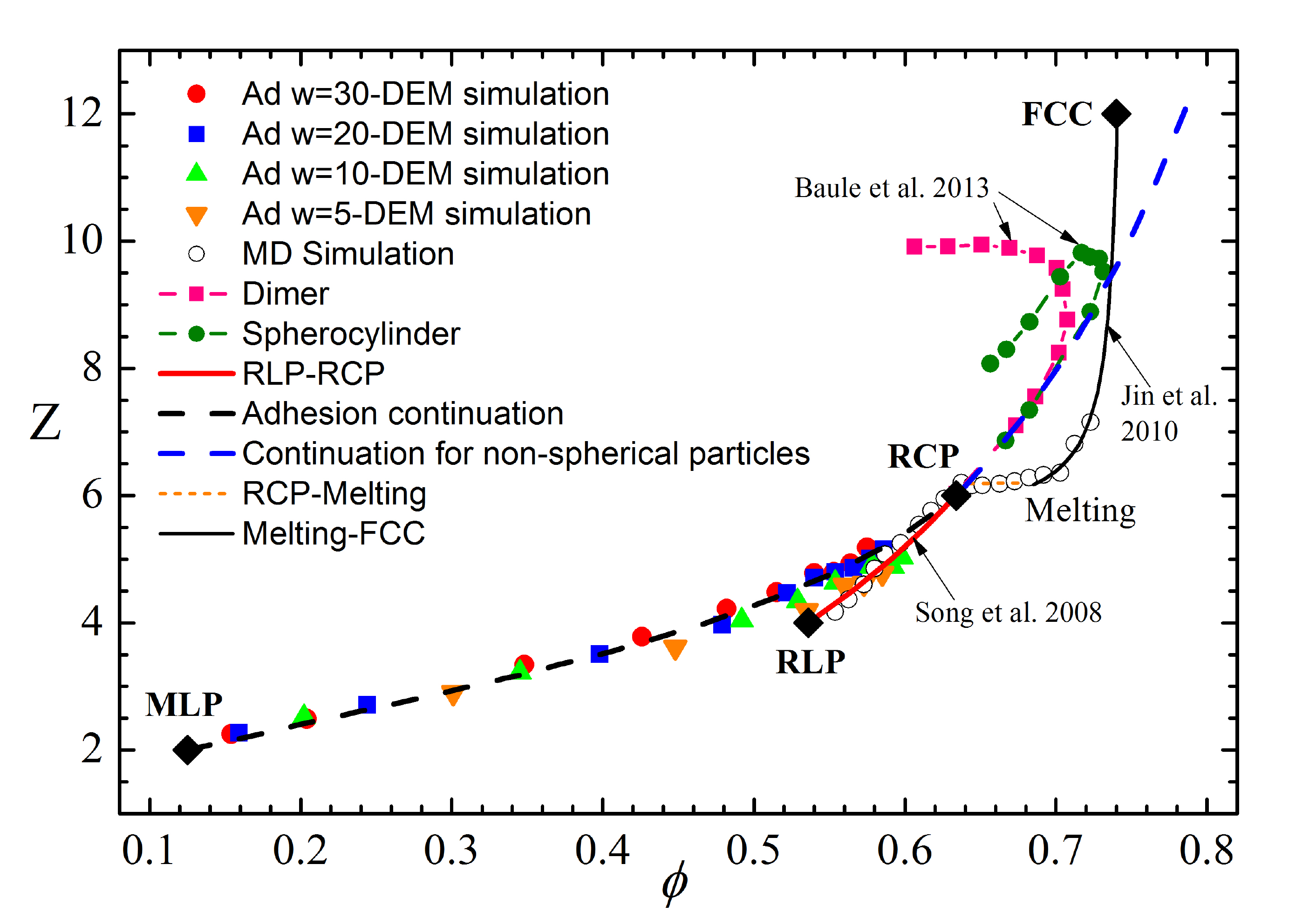}
\caption{\label{Fig_diagram}(Colors online) The phase diagram of packings of different classes of monodisperse particles including the adhesive packings. The equation of state for non-spherical particles is smoothly continued at RCP into either the adhesive or the frictional branch.
}
\end{center}
\end{figure}

Including previous results from Refs.~\cite{Song08,Jin10,Baule13} in
the $Z$-$\phi$ plane leads to a phase diagram of packings of
frictionless, frictional, and adhesive spheres, as well as
non-spherical particles (see Fig.~\ref{Fig_diagram}). The collection
of these results highlights the prominent role of the frictionless RCP
point in the phase-diagram, which appears as a focal point of the
different packing classes. We observe in particular that the equation
of state of disordered non-spherical packings is essentially smoothly
continued at RCP into either the adhesive branch or the frictional
branch. By contrast, the coexistence line from RCP to the melting
point of crystalline packings, conjectured in Ref.~\cite{Jin10}, does
not connect smoothly to any of these branches. The precise meaning of
this crucial distinction is not entirely clear to us. It suggests that
particle deformation (which parameterizes the non-spherical branch) is
a ``natural" way to increase packing densities in disordered
arrangements. On the other hand, introducing order is a more drastic
modification, similar to a distinction between discontinuous 1st and
continuous higher-order phase transitions.

In summary, we have identified a universal packing regime of adhesive small particles across $1$ to $10^2$ microns, using both DEM simulations and a statistical mechanical framework. We have shown that an equation of state for adhesive loose packings can be derived connecting the frictionless RCP with a conjectured universal MLP point in the phase diagram. The picture that emerges is that different classes of disordered packings are connected smoothly via RCP, while partially ordered phases are not. Clearly, further investigations are needed to understand the nature of packings in the vicinity of RCP, e.g., by probing the rheological properties of these packings close to jamming or by considering non-spherical adhesive particles. The MLP also deserves further attention. Open questions concern, e.g., the response of these packings to local perturbations or shear stresses. It would be highly interesting to find out whether a MLP is indeed observed in a real physical system.

\begin{acknowledgements}
The research is supported by the National Natural Science Funds of
China (No.50976058), the National Key Basic Research and Development
Program (No.2013CB228506), the basic-science funds from China Academy
of Space Technology, NSF-CMMT and DOE Geosciences Division. We
acknowledge M. Denn, J. Morris, C.~S. O'Hern, E. Brown for helpful
discussions. Thanks are due to J. Marshall, Q. Yao, G. Q.  Liu,
M. M. Yang and S. Chen for collaborations. AB acknowledges funding
under EPSRC grant EP/L020955/1.
\end{acknowledgements}

\end{document}